\documentclass[12pt]{iopart}

\usepackage{graphicx}
\begin{document}

\title[Dark Matter-Baryonic Matter Radial Acceleration Relationship]{Dark Matter-Baryonic Matter Radial Acceleration Relationship in Conservation Group Geometry }

\author{Edward Lee Green}

\address{University of North Georgia, Dahlonega, GA 30597}
\ead{egreen@ung.edu}
\begin{abstract}
Pandres has developed a theory which extends the geometrical structure of a real four-dimensional space-time via a field of orthonormal tetrads with an enlarged covariance group.  This new group, called the conservation group, contains the group of diffeomorphisms as a proper subgroup and we hypothesize that it is the foundational group for quantum geometry.  Using the curvature vector, $C_\mu$, we find a free-field Lagrangian density $C^\mu C_\mu \sqrt{-g}\,$.  When massive objects are present a source term is added to this Lagrangian density.   Spherically symmetric solutions for both the free field and the field with sources have been derived.  The field equations require nonzero stress-energy tensors in regions where no source is present and thus may bring in dark matter and dark energy in a natural way.  A simple model for a galaxy is given which satisfies our field equations.  This model includes flat rotation curves.  In this paper we compare our results with recently reported results of McGaugh, Lelli and Schombert which exhibit a new law between the observed radial acceleration and the baryonic radial acceleration.  We find a slightly different model which relates these accelerations.  In conjunction with our model, the McGaugh, Lelli and Schombert relation imply a new critical baryonic acceleration.  When applied to bulge-dominated galaxies, this critical baryonic acceleration may be used to predict the radial velocity curve value by using the radius of the bulge.
\end{abstract}

\submitto{  }
\maketitle

\section{Introduction}

Let ${\cal X}^4$ be a 4-dimensional space with orthonormal tetrad $h^{i}_{\,\,\mu}$.  Then a metric
$g_{\mu \nu}$ may be defined on ${\cal X}^4$ by $g_{\mu\nu}=\eta_{ij}\, h^i_{\,\,\mu}h^j_{\,\,\nu}$ where $\eta_{ij} =
diag\bigl\{-1,1,1,1\bigr\}$.  Whereas Einstein extended special relativity to general relativity by extending the group of transformations from the Lorentz group to the group of diffeomorphisms \cite{Ein49}, we further extend by finding the largest group of transformations for which the wave equation, $\Psi^\alpha_{\;\; ; \alpha} = 0$, is covariant.   This is the guiding principle for our theory.

A conservation law of the form $\tilde{V}^\alpha_{\;\; , \alpha}=0$, where $\tilde{V}^\alpha$ is a vector density of weight $+1$, is invariant under all transformations satisfying
\begin{equation}x^\nu_{\;\; ,\overline{\alpha}}\bigl(x^{\overline{\alpha}}_{\;\; ,\nu ,\mu} -
x^{\overline{\alpha}}_{\;\; ,\mu , \nu} \bigr) = 0 \quad . \label{e1} \end{equation}
This defines the {\it group of conservative transformations} and we easily see that the group of diffeomorphisms is a proper subgroup \cite{Pand81,Pand84,Pand09,Green09,PG03}.  Since the wave equation may be written as $\tilde{V}^\alpha_{\;\; ,\alpha} = 0 $ with $\tilde{V}^\alpha = \sqrt{- g\,}\, \Psi^\alpha$, we see that the conservation group is "the largest group of coordinate transformations under which the equation for the propagation of light is covariant"[3].  The conservation group shows potential for unifying the fields of nature \cite{Pand81}.

The geometrical content of the theory based on the conservation group is determined by the {\it curvature vector} defined by $C_\alpha \equiv h_i^{\,\,\nu}\bigl(h^i_{\,\, \alpha ,\nu}-h^i_{\,\, \nu ,\alpha} \bigr) \; = \gamma^\mu_{\;\;\alpha\mu}$, where the Ricci rotation coefficient is given by $\gamma^i_{\;\;\mu\nu}=h^i_{\;\mu ;\nu}$ [2-6]. It has been shown \cite{Pand81} that $C_\alpha$ is covariant under transformations from $x^\mu$ to $x^{\overline{\mu}}$ if and only if the transformation is conservative and thus satisfies (\ref{e1}).  A suitable scalar Lagrangian for the free field is given by
\begin{equation} {\cal L}_f= \frac1{16\pi}\int C^\alpha C_\alpha \, h \; d^4 x  \label{e2} \end{equation} where $h = \sqrt{-g}$ is the determinant of the tetrad.

If a first observer uses $x^\mu$ as coordinates and a second observer uses $x^{\bar{\mu}}$, and if $x^{\bar{\mu}}_{\; ,\nu}$ is non-diffeomorphic but is conservative, satisfying (\ref{e1}), then $x^{\bar{\mu}}$ may be interpreted as anholonomic coordinates for the first observer \cite{Scho}.  The transformation from $x^\mu$ to $x^{\bar{\mu}}$ may also be viewed as a transformation from one manifold to a second manifold.  This second manifold has a different metric and a different curvature tensor $R^\alpha_{\;\;\beta\mu\nu}$, but we argue that the masses of particles are determined by $C^\mu C^\mu$ \cite{Green11} and hence the masses of classical particles are unaltered.  When manifolds ${\cal M}_1$ and ${\cal M}_2$ are related by such a conservative transformation, we say they are in the same quantum family of manifolds.

Using $h^i_{\, \mu}=h^I_{\, \mu}\Lambda^i_I$, we have extended the field variables \cite{Green09} to include the tetrad
$h^I_{\;\mu}$ and 4 internal vectors $\Lambda^i_I$, with internal space variable $x^I$. The internal space is Lorentzian, i.e., $g_{IJ}= \eta_{IJ}\equiv diag(-1,1,1,1)$.      The definition of the Ricci rotation coefficient is also extended using the $\Lambda^i_I$ to
\begin{equation}\Upsilon^\alpha_{\;\;\mu\nu}\equiv h_I^{\;\alpha}h^I_{\;\mu
;\nu}+h_i^{\;\alpha}h^I_{\;\mu}\Lambda^i_{I,\nu} \label{e3} \end{equation} and the definition of $C_\alpha$ is
also extended to $C_\alpha \equiv \Upsilon^\mu_{\;\;\alpha\mu}$. Using these extended
Ricci rotation coefficients, one finds that
\begin{equation}C^\alpha C_\alpha = R + \Upsilon^{\alpha \beta \nu} \Upsilon_{\alpha \nu \beta}-2C^\alpha_{\; ;\alpha}
- \eta^{ij}h_j^{\;\nu}h_I^{\;\alpha}(\Lambda^I_{i,\alpha ,\nu}-\Lambda^I_{i,\nu
,\alpha})\quad , \label{e4} \end{equation}
where $R$ is the usual Ricci scalar curvature.  Comparing (\ref{e4}) with GR we see that the Lagrangian density of the free field contains additional terms \cite{PG03,Green11}.

Setting $\delta {\mathcal L}_f \, = 0$ leads to field equations.  The free-field Lagrangian does not depend directly on the field variables $L^i_I$ and $h^I_{\;\alpha}$ but only on the regular tetrad $h^i_{\;\alpha} \, = \; L^i_I \, h^I_{\;\alpha}$.  The field equations (see Pandres \cite{Pand81,Pand84,Pand09,PG03} ) are:
\begin{equation} C_{\mu ;\nu} - C_{\alpha}\Upsilon^\alpha_{\;\mu\nu} - g_{\mu\nu}C^\alpha_{\; ;\alpha} - \frac12 C^\alpha C_\alpha \; = \;
0  \label{fe0} \end{equation}

We assume for the remainder of this paper that $\Lambda^I_i =\delta^I_i $ (i.e., no internal fields - only gravity). In this case, an identity for  the Einstein tensor is
\begin{eqnarray}  G_{\mu\nu}= & C_{\mu ; \nu}- C_\alpha \Upsilon^\alpha_{\; \mu\nu} -g_{\mu\nu}C^\alpha_{\; ;\alpha}-\frac12 g_{\mu\nu}C^\alpha C_\alpha   \nonumber \\ \quad & +\Upsilon^{\;\;\alpha}_{\mu \;\; \nu ;\alpha}+\Upsilon^\alpha_{\;\;\sigma
\nu} \Upsilon^\sigma_{\;\; \mu \alpha} + \frac12 g_{\mu\nu}\Upsilon^{\alpha \beta
\sigma} \Upsilon_{\alpha \sigma \beta}  \nonumber \end{eqnarray}   Using (\ref{fe0}) we see that the field equations may be also expressed in the form
\begin{equation}G_{\mu\nu} = \Upsilon^{\;\;\alpha}_{(\mu \;\; \nu) ;\alpha}+  \Upsilon^\alpha_{\;\;\sigma (\nu} \Upsilon^\sigma_{\;\; \mu) \alpha}  + \frac12 g_{\mu\nu}\Upsilon^{\alpha \beta
\sigma} \Upsilon_{\alpha \sigma \beta}\;  \equiv \; \; 8\pi \bigl(\mathbf{T}_{\rm f}\bigr)_{\mu\nu}  \label{e9} \end{equation}
with free-field stress energy tensor $\mathbf{T}_{\rm f}$. The additional terms evident in $\mathbf{T}_{\rm f}$ suggest that this new geometry could bring in dark matter and dark energy in a natural way \cite{Green11}.

\par \phantom{D} \par

\section{Spherically symmetric solutions.}

Spherically symmetric solutions (using labels $(t,r,\theta,\phi)$) of the field equations for a free field (\ref{fe0}) are given by the tetrad
\begin{equation} h^i_{\;\; \mu} = \left[  \begin{array}{cccc}
\; e^{\Phi} & 0 & 0 & 0 \\
0 &\; \bigl(1+\frac12 r\Phi^\prime \bigr)\sin \theta \cos \phi \; & \; r\cos\theta\cos\phi \; & \; -r\sin\theta\sin\phi \\
0 & \bigl(1+\frac12 r\Phi^\prime \bigr)\sin\theta\sin\phi & r\cos\theta\sin\phi & \;\; r\sin\theta\cos\phi \\
0 & \bigl(1+\frac12 r\Phi^\prime \bigr)\cos\theta\qquad & -r\sin\theta\qquad & 0
\end{array}
\right]
\label{e16} \end{equation} for which $C_\mu=0$ and hence (\ref{fe0}) is satisfied \cite{Green11}.  (The upper index refers to the row and the prime indicates differentiation with respect to $r$.)  The metric is
\begin{equation}ds^2= -e^{2\Phi(r)} dt^2 +  \bigl(1+\frac12 r\Phi^\prime(r) \bigr)^2dr^2+r^2d\theta^2
+r^2\sin^2\theta d\phi^2 \quad . \label{e17} \end{equation}
The Einstein tensor is diagonal and the nonzero components are (with $\Phi$ representing $\Phi(r)$)
\begin{equation}
G_{tt} =\; \frac{e^{2\Phi}\biggl(
\frac18(r\Phi^\prime)^3+\frac34(r\Phi^\prime)^2+2r\Phi^\prime+r^2\Phi^{\prime\prime}\biggr)
}{r^2\biggl(1+\frac12r\Phi^\prime\biggr)^3} \quad ,  \label{e18}
\end{equation}
\begin{equation}
G_{rr}=\; \frac{r\Phi^\prime - \frac14(r\Phi^\prime)^2}{r^2} \label{e19}
\end{equation} and
\begin{equation}
\frac{\;\; G_{\theta\theta} \;\;}{\;\; r^2\;} \; = \; \frac{G_{\phi\phi}}{r^2\sin^2\theta} \;  =  \; \frac{\; \frac12(r\Phi^\prime)^3+(r\Phi^\prime)^2
+\frac12 r\Phi^\prime+\frac12r^2\Phi^{\prime\prime}}{r^2\bigl(1+\frac12r\Phi^\prime \bigr)^3}  \label{e20}
\end{equation}
As $G_{\mu\nu}$ is generally nonzero, we see that these free-field solutions may automatically include dark matter and dark energy.

Let
\begin{equation}
  m(r)\, \equiv \, r\, - \frac{r}{(1+\frac12r\Phi^\prime)^2} \label{wDef}
\end{equation}
Thus $  \Phi^\prime(r)= \frac2r\biggl[\bigl(1-\frac{m(r)}{r}\bigr)^{-\frac12} - 1\biggr] \, . $  Notice that when $m(r)<< r$ then $\Phi^\prime(r) \approx \frac{m(r)}{r^2}$ (this will be used in the Section 4 below).  Let $G^t_t = - 8\pi \rho_f$, where $\rho_f$ is the density mass-energy of the free-field (see \cite{MTW} or \cite{Wein72}).  Using (\ref{e18}) we find that \cite{Green11}
\begin{equation}  \; 8\pi \rho_f  =  \frac1{r^2} m^\prime(r)   \label{e21} \end{equation}
We now identify the function $m(r)$ as the total mass inside a ball of radius $r$.   Also
\begin{equation}g_{rr}=\bigl(1+\frac12r\Phi^\prime\bigr)^2=\biggl(1-\frac{m(r)}{r}\biggr)^{-1} \; \; , \label{e23} \end{equation} and
\begin{equation}g_{tt}= -e^{2\Phi(r)} \quad \mathtt{\rm , where } \;\;   \Phi(r)= \int \frac2r\biggl[\Bigl(1-\frac{m(r)}{r}\Bigr)^{-\frac12} - 1\biggr] \, dr \quad \label{e24} \end{equation}  (this defines $\Phi(r)$ up to a constant).

The radial pressure denoted by $\, p_R$ is determined by $G^r_r = 8\pi p_R$. In terms of $m(r)$ the result is
\begin{equation}
8\pi p_R = \frac{ 4r\sqrt{1-\frac{m(r)}{r}} - 4r + 3m(r)}{r^3} \quad . \label{e26}
\end{equation}
The tangential pressure denoted by $p_T$ is determined by $8\pi p_T= G^\theta_\theta =G^\phi_\phi$ and in terms of $m(r)$ we find that
\begin{equation}
8\pi p_T=\frac{8r-\frac92 m(r)-8r\sqrt{1-\frac{m(r)}{r}}+ \frac12 r m^\prime(r)}{r^3} \quad . \label{e28}
\end{equation}
Since $p_R\neq p_T$ shear stresses are present and we see that $\bigl(T_{\rm f}\bigr)_{\mu\nu}$ does not model a perfect fluid.  The conservation of energy condition, $T^\mu_{\;\;\nu ;\mu}=0$ is trivially satisfied for $\nu=0, 2 $ and $3$, but when $\nu=1$ we find that
\begin{equation}
\bigl(\, \rho + p_R \bigr)\Phi^\prime \; = \, - p_R^{\; \prime} + \frac2r \, \bigl(p_T-p_R\bigr)  \label{e29a}
\end{equation}
which indicates that particles do not move along geodesics due to pressures (see \cite{MTW}, page 601).

Using the ideal gas law, $PV=nRT$, we define the temperature per unit mass of the halo as
\begin{equation}
T \equiv \; \frac{\bar{p}}{\rho} \; \, = \; \frac13 + \frac{2\Bigl(\; 1-\sqrt{1-\frac{m(r)}r }\;\Bigr)^2}{m^\prime (r)}   \label{e30}
\end{equation}
with $m(r)$ given by (\ref{wDef}) and with the average pressure defined by $\bar{p}=(p_R+p_T+p_T)/3$.

In order for a free-field, spherically-symmetric solution to agree with the weak-field solution as $r\to \infty$, we will require that $\lim_{r\to\infty} m(r)\,=\,M$, where $M$ is the total mass of the star as measured for very large values of $r$ (this includes dark matter also).  Furthermore we assume that $m(r)$ is a non-decreasing, piecewise differentiable function of $r$.  We also denote the minimum value of $r$ for which the free-field solution is defined by $R_B$ ($B$ indicates the baryonic radius).  The value of $r\,=\, R_B$ may correspond to the radius of the bulge or galactic center.  Our theory does not uniquely determine the $m(r)$ function according to our assumptions given above.  With the temperature determined by (\ref{e30}) we seek isothermal haloes which would be in thermodynamic equilibrium.  We use the isothermal concept which is energy concept to select the classical solution from the family of manifolds which are allowed by our field equations \cite{DSS}.

\section{Free-field Model for Galaxies. }

We have found two models which satisfy our field equations and conditions on $w(r)$ which are isothermal.  One of them reasonably models the dark matter component of galaxies.  In the following we assume that the field is non-free (sources present) for $0\leq r \leq R_B$ where ordinary baryonic matter is predominant, in the bulge or galactic center.    For $r>R_B$ we assume that there are no sources or that the sources are negligible.  Let $m(r)$ be a function which gives the total mass-energy within a sphere of radius $r$.  Let $\lim_{r\to\infty}m(r)=M$ be the total asymptotic mass.  At the surface of the bulge or galactic center, the value, $m(R_B)=M_{B}$, represents nearly all of the baryonic mass of the galaxy.  Although $M_B$ is not completely determined by the value at $R_B$, we assume that the difference is small in relative terms.  Numerical solutions may be utilized to more closely model actual galaxies, but our simplified model will nevertheless lead to several important results.

Define a dimensionless  positive constant $k$ as
\begin{equation}
  k=\frac{\, M_B}{R_B}  \label{kdef}
\end{equation}  Using (\ref{e30}), we note that the linear function $\, m(r)= \,  k r \,$ results in a constant temperature per unit mass of $T=\frac13 + \frac{2\Bigl( \; 1- \sqrt{1-k} \;\Bigr)^2}{k} \approx \frac13 + \frac{k}2$ when $k<< 1$.
Thus we define
\begin{equation}
m(r)= \Biggl\{
\begin{array}{cc}
 \, k \, r \quad &  ,  R_B\leq r \leq R  \\
M  \quad & , \; r > R \qquad
 \end{array}
\label{e63} \end{equation}
where $R$ is defined by $\, R\, \equiv \frac{\,R_B M}{M_B}$.  (Note:  $\; \frac{M}{M_B}\, = \, \frac{R}{R_B}\,$.)
According to this model, the halo extends to $r\,=\,R\,$.  As a typical example, $\frac{\, M\,}{M_B} = 10$ which leads to a halo radius of $R=10R_B$.    We note that this is a free-field model, with the curvature vector vanishing ( $C_\mu =0 \,$).

For this model the line element is (for $R_B\leq r \leq \, R \,$)
\begin{equation} ds^2 = -C r^{\frac{\, 4(1-\sqrt{1-k})}{\sqrt{1-k}}} dt^2 + \Bigl( 1-k \Bigr)^{-1} dr^2  + r^2 (d\theta^2 + \sin^2\theta d\phi^2)  \label{e64} \end{equation}
where $C$ is a constant chosen so that the metric will match the metric of (\ref{e53})  when $\, r=\,R\,$.
The density and pressures are (for $R_B\leq r \leq \, R\,$)
\begin{eqnarray} 8\pi\rho &=  \frac{k}{ r^2}            \nonumber \\
8\pi p_r &= \frac1{r^2}\Bigl(1-\sqrt{1-k}\Bigr)\Bigl(3\sqrt{1-k}-1\Bigr) \label{e65} \\
8\pi p_T &= \frac4{r^2}\Bigl(1-\sqrt{1-k}\Bigr)^2 \nonumber \end{eqnarray}
For $r>\, R\, $, (\ref{e63}) indicates that $m(r)= M$.  From (\ref{wDef}) we have $M \,= \, r \, -\frac{r}{(1+\frac12r\Phi^\prime)^2}$ and hence
\begin{equation}\Phi(r) = \int \Biggl[\; \frac{2}{r\sqrt{1-\frac{M}r}}-\frac{2}{r} \, \Biggr] \; dr \label{e50} \end{equation}
which can be easily integrated.  The value of the arbitrary constant is determined by the condition that $g_{tt}\approx -1 + \frac{2M}{r}$ for large $r$.  The result
\begin{equation}  g_{tt}=-\frac1{256}\biggl(1+\sqrt{1-\frac{M}r} \biggr)^8 \quad .  \label{e52} \end{equation}
We thus obtain the following line element for $r>\,R\, \;$:
\begin{equation}ds^2 = -\frac1{256}\biggl(1+\sqrt{1-\frac{M}r}\biggr)^8 dt^2 + \Bigl(1-\frac{M}r\Bigr)^{-1} dr^2 + r^2 d\theta^2 + r^2\sin^2\theta d\phi^2 \label{e53} \end{equation}
The constant of (\ref{e64}) can now be determined and is found to be
\begin{equation}
  C \, = \, \frac{(1+\sqrt{1-k})^8}{256} \cdot \Bigl(R \Bigr)^{\frac{4(\sqrt{1-k}-1)}{\sqrt{1-k}}}
\end{equation}
and thus for $R_B \leq r \leq R$:
\begin{equation}
  ds^2 = - \frac{(1+\sqrt{1-k})^8}{256} \Bigl( \frac{r}{R} \Bigr)^{\frac{\, 4(1-\sqrt{1-k})}{\sqrt{1-k}}} dt^2 + \frac{dr^2}{1-k}   + r^2 (d\theta^2 + \sin^2\theta d\phi^2)
\end{equation}
Using (\ref{e18}-\ref{e20}), we find that where $r>\, R\,$ the density is zero, but the pressures are nonzero:
\begin{eqnarray}  8\pi \rho \; &= \; 0 \qquad \qquad \qquad  \nonumber \\
8\pi p_R &= \; \frac{M\Bigl(3\sqrt{1-\frac{M}r}-1 \Bigr)}{r^3\Bigl(1+\sqrt{1-\frac{M}r}\Bigr)} \qquad \qquad , \quad  r>\, R  \label{e56} \\
8\pi p_T &= \; \frac{-M\Bigl(9\sqrt{1-\frac{M}r}-7\Bigr)}{2r^2\Bigl(1+\sqrt{1-\frac{M}r}\Bigr)}
\nonumber \end{eqnarray}

\subsection{Equations of Motion of Test Bodies.}

An efficient procedure for finding equations of motion is one that extremalizes an appropriate Lagrangian.  We follow de Felice and Clarke \cite{dFC}       with a Lagrangian for a particle in a field with nonzero stress energy tensor.  Specifically,  we introduce a source term in the Lagrangian density $L=C^\mu C_\mu + L_s$:
\begin{equation}  L_s = \rho(x)  = \mu \int \delta_\epsilon^4(x-\gamma(s))(-u^\mu u_\mu)^{\frac12} ds   \label{e67} \end{equation}
where $\delta_\epsilon^4$ approximates the Dirac delta function with a space-like volume of $\epsilon$ which yields the usual Dirac delta function in the limit as $\epsilon\to 0$.  The path of the particle is given by $\gamma(s)$ and its velocity is $u^\mu=\frac{dx^\mu}{d\tau}$.  We will use the "dot" for the components of $u^\alpha$, i.e. $u^\alpha = \langle \dot{t},\dot{r},\dot{\theta},\dot{\phi}\rangle$.  Let $\mu$ denote the mass of the particle. The condition  $T^{\beta \alpha}_{\; \;\;\; ; \beta }=0$ leads to \cite{Green11}
\begin{equation}  \frac{\mu}{\sqrt{-u^\nu u_\nu}} \, u^\beta u^\alpha_{\; ;\beta} = \; \delta^\alpha_1
F_p   \;    \label{e68} \end{equation}
From (\ref{e29a}) and [14], we see that $F_p \equiv \epsilon \, g^{rr} \, \bigl(-p_R^{\; \prime} + \frac2r \, \bigl(p_T-p_R\bigr)\,\bigr) $. Thus from (\ref{e65}) we find that $F_p = \frac{\epsilon (1-k)(1-\sqrt{1-k})^2}{\pi \, r^3}$ and so we may express (\ref{e68}) as
\begin{equation}
   \frac1{\sqrt{-u^\nu u_\nu}} \, u^\beta u^\alpha_{\; ;\beta} = \; \delta^\alpha_1 \, \frac{\, (1-k)(1-\sqrt{1-k})^2}{\pi \, \tilde{\rho} \, r^3}  \label{e68a}
\end{equation}
where $\tilde{\rho}=\frac{\mu}{\epsilon}$ is the average density of the test body.
When $\alpha\neq 1$, (\ref{e68a}) implies  $u^\beta u^\alpha_{\; ;\beta} = 0$ which is the usual geodesic equation.
We choose to restrict the motion to the $\theta=\frac{\pi}2$ plane.  Using $L$ for the constant of angular momentum and $E$ for the energy of the test body, one finds that (for $\theta\equiv \frac{\pi}2$)
\begin{equation}
  \dot{\phi} = \frac{L}{r^2} \label{phidot}
\end{equation}
and
\begin{equation}
  \dot{t} =  \frac{256\, E}{(1+\sqrt{1-k})^8} \cdot \biggl(\frac{r}{R} \biggr)^{\frac{4(\sqrt{1-k}-1)}{\sqrt{1-k}}}  \qquad , \; R_B \leq r \leq \, R \label{tdot}
\end{equation}

For the metric (\ref{e64}) (with $\theta \equiv \frac{\pi}2$)
\begin{equation}  \ddot{r} + \frac{k(1+\sqrt{1-k})^7\sqrt{1-k} \,}{128 r} \biggl(\frac{r}{R} \biggr)^{\frac{4(1-\sqrt{1-k})}{\sqrt{1-k}} } \, \dot{t}^2 \,  -  r \,(1-k)  \dot{\phi}^2  =  \frac{\sqrt{-u^\nu u_\nu}\, F_p}{\mu}  \label{rddot}
\end{equation}
using (\ref{phidot}) and (\ref{tdot}) and the normalization $u^\nu u_\nu = -1$, we get
\begin{eqnarray}  \ddot{r} \, + \, \frac2r \biggl(\frac{1-\sqrt{1-k}}{\sqrt{1-k}} \biggr) \, \dot{r}^2 \, + \; \frac2r \Bigl(\sqrt{1-k}+k-1\Bigr) \nonumber  \\ \qquad\qquad\qquad + \;  \frac{L^2}{r^3} \Bigl( 2\sqrt{1-k} + 3k -3 \Bigr) \,  = \; \frac{(1-k)(1-\sqrt{1-k})^2}{\pi\tilde{\rho}r^3}   \label{rddot2}
\end{eqnarray}

For pure radial motion ($L=0$), (\ref{rddot2}) with $R_B< r < R$ yields  \begin{equation} \ddot{r} = \; -\frac{2}{r}\Bigl(\sqrt{1-k}+k-1)-\frac2r\biggl(\frac{1-\sqrt{1-k}}{\sqrt{1-k}}\biggr)\dot{r}^2 + \frac{(1-k)(1-\sqrt{1-k})^2}{\pi \tilde{\rho} r^3}   \label{e85} \end{equation}
If we again assume that $k$ is small, then the $F_p$ term on the right may be dropped.  Objects with low radial velocities experience an acceleration of
\begin{equation} \ddot{r} \approx - \frac{k}{r} \, + \, \frac{k^2}{r} \; \qquad \mathtt{\rm , \;\; for } \quad R_B \; \; \leq r \leq R \label{e86} \end{equation}
to second order in $k$.  Thus to first order,
\begin{equation}
  \ddot{r} \; \approx \; - \frac{k }{r } \qquad , \mathtt{\rm \;\; for } \quad R_B \; \; \leq r \leq R \label{e86a}
\end{equation}

\subsubsection{Flat Rotation Velocity Curves and the Baryonic Tully-Fisher Relation.}
Assuming that $k$ is very small (typically less than $10^{-2}$) we find that (to first order in $k$) $\;g_{tt} \approx -1$.  If $\tilde{\rho}$ is very small, the $\frac{F_p}{\mu} \approx \frac{k^2}{4\pi \tilde{\rho} r^3}$  term could be significant, but for typical galaxies this is not the case.  This is because the values of $r$ are typically on the order of $10^{22}$ cm.  Thus the $F_p$ term is estimated to be much, much less than $k^2$ and we will drop it in our second order approximations. To compute $v_r$, we use the formula, $v_r = r\frac{d\phi}{d t} = \frac{r\dot{\phi}}{\dot{t}}$ and use (\ref{rddot}) setting $\dot{r}$ and $\ddot{r}$ to zero for a circular orbit.   The result to second order is
\begin{equation}  (v_r)^2 \; \approx \; k  \, + \Bigl(  \frac{2r}{R} -\frac{13}4 \Bigr) k^2 \quad , \;\; \mathtt{\rm for } \quad R_B < r \leq R  \label{e84a} \end{equation}
The second order differences in radial acceleration  between (\ref{e86}) and (\ref{e84a}) may be explained by the nonzero pressure terms in the stress-energy tensor (\ref{e65}).  To first order,
\begin{equation}
(v_r)^2 \approx k \qquad , \mathtt{ \rm \; i.e.} \quad  v_r \approx \sqrt{k} \quad , \;\; \mathtt{\rm for } \quad R_B < r \leq R \label{e84} \end{equation}
For example if $k = 10^{-6}$ (fairly typical for a galaxy) then $v_r\approx 10^{-3} \approx 300\, $km/s.  This constant velocity curve would extend from the disk out to the boundary of the halo.

The baryonic Tully-Fisher relation \cite{TF} states that $M_B \propto (v_r)^4 $.  From (\ref{kdef}) and (\ref{e84}) we see that $M_B \propto k^2 = \Bigl(\frac{M_B}{R_B}\Bigr)^2$.  Thus we have a general result:
\begin{equation}
  \frac{M_B}{(R_B)^2} \propto 1 \; \Rightarrow \; \frac{M_B}{(R_B)^2 } \, = \; \biggl(\frac{e}{e-1}\biggr)^2 g_\dagger \, \approx  \; 2.50 g_\dagger  \label{gdagger}
\end{equation}
where $g_\dagger$ is a universal constant acceleration and the coefficient, $\bigl(\frac{e}{e-1}\bigr)^2$ is chosen for convenience.

\section{Implications of our Free-field Galaxy Model and the Radial Acceleration Relation.}

McGaugh, Lelli and Schombert \cite{McGL} have recently reported a correlation between the radial acceleration that is observed and the radial acceleration due to the baryons.  From the definition of $k$, (\ref{kdef}) we see that $k$ is basically the gravitational potential at $r=R_B$ and in our model it also is basically the gravitational potential at $r=R$, since $\frac{m(R)}R = k$.  Using the value of $M_B$, we have the baryonic acceleration given by $g_B= \frac{M_B}{r^2} = \frac{kR_B}{r^2}$.  Thus $r = \sqrt{\frac{kR_B}{g_B}}$.  We relate the observed acceleration to the rotation curve via $\; \frac{ v_r^{\; 2}}{r} = g_{obs}$. We assume that $k$ is very small and use the first order approximation (\ref{e84}) to find $g_{obs} \approx \; \frac{k}r  $, for $\, R_B \leq r \leq R$. Thus our theory leads to the following relation (to first order in $k$):
\begin{equation}
  g_{obs} = \sqrt{\frac{k}{R_B}}\sqrt{g_B} \qquad \mathtt{\rm  for } \quad  R_B \leq r \leq R \label{rar1}
\end{equation} According to \cite{McGL},
\begin{equation}
  g_{obs} \, = \;  \frac{g_B}{\, 1-e^{-\sqrt{g_B /g_\dagger}}}  \label{rar0}
\end{equation}
is a universal radial acceleration relation for galaxies.  McGaugh, Lelli and Schombert \cite{McGL} based this on analysis of 153 galaxies, including bulge-dominated spirals, disk-dominated spirals and gas-dominated dwarfs. They find that the value of the acceleration-parameter $g_\dagger$, which accurately models the relationship given in (\ref{rar0}), is approximately $\, 1.20 \times 10^{-10} $ m/s$^2$.  We stipulate that the two formulae (\ref{rar1}) and (\ref{rar0}) should agree when when $g_B = g_\dagger $.  Thus we find
\begin{equation}
  \frac{e}{e-1} \sqrt{g_\dagger} = \sqrt{\frac{k}{R_B}}  \quad \Longleftrightarrow \quad \biggl(\frac{e}{e-1}\biggr)^2 g_\dagger = \frac{M_B}{(R_B)^2}  \label{rar2}
\end{equation}
agreeing with (\ref{gdagger}).  Thus our theory suggests that
\begin{equation}
  g_{obs} \, = \; \frac{e}{e-1} \sqrt{\, g_\dagger \, g_B }  \label{rar3}
\end{equation}
When $g_{obs} = g_B$, the implication is that there is no significant amount of dark matter.  From (\ref{rar3}) we find that
\begin{equation}
  g_{obs} \, = g_B\quad \Rightarrow \quad \sqrt{g_B} = \frac{e}{e-1} \sqrt{g_\dagger} \; \equiv \;  \sqrt{ g_{B_c}} \label{rar4}
\end{equation}
which we call the critical baryonic-acceleration constant denoted by $g_{B_c}$.  Hence from the value found for $g_\dagger$ (see \cite{McGL}), we find the critical baryonic-acceleration constant to be
\begin{equation}
  g_{B_c} \approx \; 3.0 \times 10^{-10} \; \mathtt{ \rm m/s}^2 \label{rar5}
\end{equation}
From (\ref{rar2}) and (\ref{rar4}), we also find that $g_{B_c} = \frac{k}{R_B}$ which matches well with our hypothesis that for $r<R_B$, there is only baryonic matter present.  Finally, define the parameter $\alpha > 1$ for a given galaxy, by $\alpha = \frac{R}{R_B} = \frac{M}{M_B}$, which indicates the ratio of the total mass to the baryonic mass.  The consensus value is $\alpha \approx 6$.  Then when $g_B <  \frac1{\alpha^2} g_{B_c}$, we find that $g_{obs} = \alpha g_B$.    Thus we propose the following model for the relationship between the baryonic acceleration  and the observed acceleration:
\begin{equation}
  g_{obs (\alpha)} =
     \left\{              \begin{array}{ll}  \;\; \alpha g_B & , \;\; \;\;\; \, g_B\leq \frac1{\alpha^2} g_{B_c} \\
            \sqrt{(g_{B_c})\, g_B \,}  & , \; \frac1{\alpha^2} g_{B_c} < g_B < g_{B_c} \\  \;\; g_B &  , \;\;\;\;\;\; g_B \geq g_{B_c}                \end{array}       \right. \label{rar6}
\end{equation}
The graph of our radial acceleration relation is shown in Figure 1 for values of $g_B$ which are greater than $\frac1{\alpha^2}g_{B_c}$.  We also include the plot of (\ref{rar0}) using the value $g_\dagger = 1.2 \times 10^{-10}$ m/s$^2$.  The difference between these graphs is small and appears to be within 1.0 rms according to McGaugh, Lelli and Schombert \cite{McGL}.  We note that our model is highly idealized and thus when numerical solutions of our field equations are produced that more closely resemble the actual distribution of baryonic matter, then we expect that the model will approach the curve given in (\ref{rar0}).

When we combine the predictions of our theory with those of McGaugh, Lelli and Schombert, the new and interesting conclusion is that $g_{B_c}\approx 3.0 \times 10^{-10}$ m/s$^2$ is a universal critical acceleration.  The acceleration at the baryonic radius, $R_B$, is given by $g_B = \frac{kR_B}{(R_B)^2} = \frac{k}{R_B} = g_{B_c}$.  If $g_{B_c}$ is considered to be the fundamental constant that governs the dividing line between baryonic and dark matter, then $k$ may be determined for a given galaxy by finding the radius at the bulge.  From (\ref{e84}) we would then have a prediction of the value of $v_r$.  However, this should only apply to bulge-dominated galaxies which seem to fit more closely with the model proposed here.

\begin{figure}
\includegraphics[width=0.8\textwidth,height=0.7\textwidth]{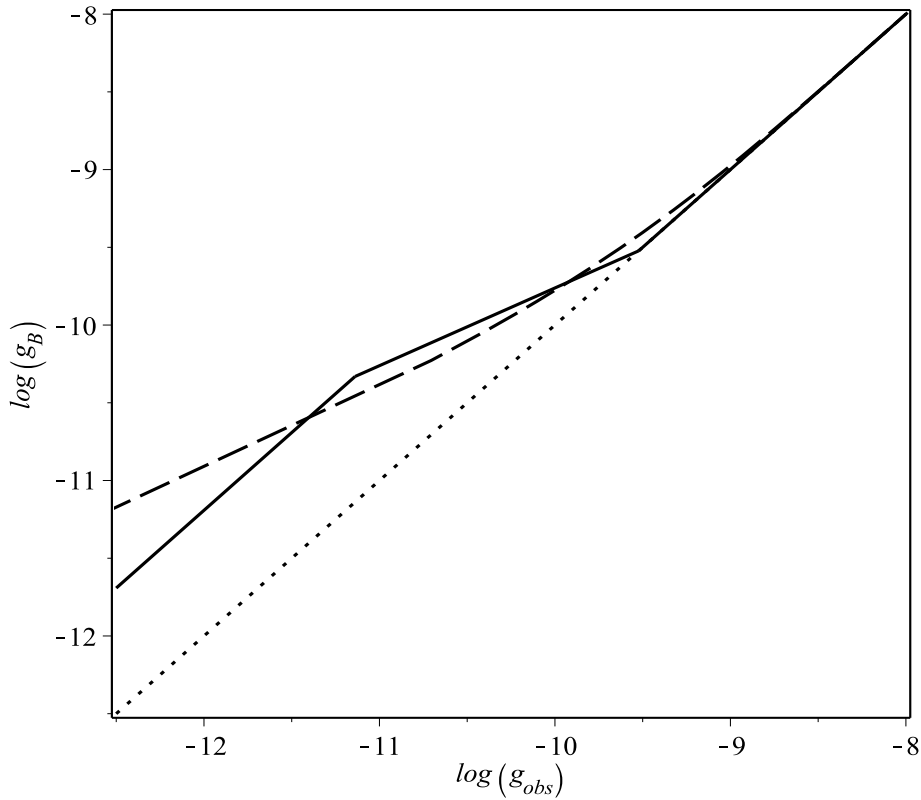}
\caption{Our radial acceleration relation (\ref{rar6}) with $\alpha= 6.43$ (solid) versus the relation found by McGaugh, Lelli and Schombert (dashed).  Also the line of unity is shown (dotted).  (accelerations measured in units of m/s$^2$)}
\label{fig:1}
\end{figure}

\par \phantom{D} \par

\section{Conclusion.}

The theory based on the conservative transformation group leads to a spherically symmetric solution for the free field, which must have nonzero densities and pressures.  Our isothermal solution reasonably models galaxies that are bulge-dominated, but also gives insights into what more accurate models based on the conservative transformation group may predict.  Namely, that along with the baryonic Tully-Fisher relationship and the radial acceleration relation of McGaugh, Lelli and Schombert, we obtain the prediction of a universal critical acceleration, $g_{B_c}$.  Further progress in developing isothermal models that more closely model disk-dominated spirals and dwarf galaxies may strengthen these results.  Our extension of the covariance group to the conservative transformation group leads to automatic inclusion of dark matter and dark energy.  As our theory suggests along with many recent results, dark matter and baryonic matter are not independent of one another.  It appears that dark matter is simply a geometrical effect of baryonic matter, but much work is needed to fully develop this theory.

\subsection*{Acknowledgments}  The author would like to thank Dave Pandres for the long years of pioneering work on this theory and for many helpful suggestions.  Also the author would like to thank Peter Musgrave, Denis Pollney and Kayll Lake for the GRTensorII software package which was very helpful.

\end{document}